%% file: aaai25.tex
\documentclass[letterpaper]{article} 

\usepackage{aaai25} 

\usepackage{times}  
\usepackage{helvet}  
\usepackage{courier}  
\usepackage[hyphens]{url}  
\usepackage{graphicx} 
\usepackage{natbib}  
\usepackage{caption} 
\usepackage{algorithm}
\usepackage{algorithmic}

\usepackage{newfloat}
\usepackage{listings}
\DeclareCaptionStyle{ruled}{labelfont=normalfont,labelsep=colon,strut=off} 

\usepackage{caption}
\usepackage{subcaption}

\floatstyle{ruled}
\newfloat{listing}{tb}{lst}{}
\floatname{listing}{Listing}
\usepackage{color,soul}

\title{To Err is AI : A Case Study Informing LLM Flaw Reporting Practices}
\author {
    Sean McGregor,\textsuperscript{\rm 1, *}
    Allyson Ettinger,\textsuperscript{\rm 2}
    Nick Judd,\textsuperscript{\rm 1}
    Paul Albee,\textsuperscript{\rm 2}
    Liwei Jiang,\textsuperscript{\rm 2}
    Kavel Rao,\textsuperscript{\rm 2}
    Will Smith,\textsuperscript{\rm 2}
    Shayne Longpre,\textsuperscript{\rm 3}
    Avijit Ghosh,\textsuperscript{\rm 4}
    Christopher Fiorelli,\textsuperscript{\rm 2}
    Michelle Hoang,\textsuperscript{\rm 5}
    Sven Cattell,\textsuperscript{\rm 6}
    Nouha Dziri\textsuperscript{\rm 2}
}
\affiliations {
    \textsuperscript{\rm 1}UL Research Institutes,
    \textsuperscript{\rm 2}Allen Institute for AI,\\
    \textsuperscript{\rm 3}Massachusetts Institute of Technology,
    \textsuperscript{\rm 4}Hugging Face,
    \textsuperscript{\rm 5}SeedAI,
    \textsuperscript{\rm 6}AI Village
}

\begin{document}

\maketitle

\begin{abstract}
In August of 2024, 495 hackers generated evaluations in an open-ended bug bounty targeting the Open Language Model (OLMo) from The Allen Institute for AI. A vendor panel staffed by representatives of OLMo's safety program adjudicated changes to OLMo's documentation and awarded cash bounties to participants who successfully demonstrated a need for public disclosure clarifying the intent, capacities, and hazards of model deployment. This paper presents a collection of lessons learned, illustrative of flaw reporting best practices intended to reduce the likelihood of incidents and produce safer large language models (LLMs). These include best practices for safety reporting processes, their artifacts, and safety program staffing.

\end{abstract}

\input{sections/0_introduction}

\input{sections/1_llm_vendor_challenges}

\input{sections/2_discussion}


\section{Acknowledgments}
\input{sections/3_acknowledgements}

\bibliography{aaai25}

\end{document}

%% file: sections/0_introduction.tex
\section{Introduction} \label{introduction}
On August 9th, 2024, the organizers of Generative Red Team 2 (GRT2) declared to
a packed session at the DEF CON hacking conference that the assembled hackers
would fail to find flaws with the Open Language Model (OLMo)
\cite{groeneveld2024olmoacceleratingsciencelanguage}. This tongue-in-cheek taunt
served as motivation for the hackers to root out cases in which the large language
model (LLM) failed to live up to the intentionally lofty and unachievable claims
made in the event’s model documentation. Over the next two days, 495 participants
prepared 200 ``flaw reports," detailing ``any unexpected model behavior that is
outside of the defined intent and scope of the model design"
\cite{cattell2024coordinatedflawdisclosureai}. Of the \$10,000 in bounty awards
in the prize pool, \$7,400 was paid out to participants.

This event addresses a burgeoning need for broader participation in the evaluation of AI systems' safety, security, and trustworthiness.
Recent work points to a growing spectrum of hazards from generative AI  \citep{kapoor2024societal,weidinger2022taxonomy,lakatos2023revealing,thiel_generative_2023,li2023chatgpt,renaud2023chatgpt,soice2023can,ftc2023voice} that bolster the case for independent and community-driven algorithmic flaw evaluations \citep{Elazari2018,kenway2022bugs,birhane2024ai} and coordinated flaw disclosure protocols designed specifically for AI \cite{cattell2024coordinatedflawdisclosureai,householder2024lessons}.
While prior work has already contributed a rich body of AI flaws \citep{yong2023low,nasr2023scalable,parrish2023adversarial,qi2023fine,kotha2023understanding}, a lack of infrastructure for responsible disclosure, or researcher protections has stifled much-needed evaluations \citep{longpre2024safeharboraievaluation}.
This paper documents a large-scale attempt to operationalize recommendations addressing these shortcomings at DEF CON 2024.

\begin{figure}
     \centering
         \centering
         \includegraphics[width=0.45\textwidth]{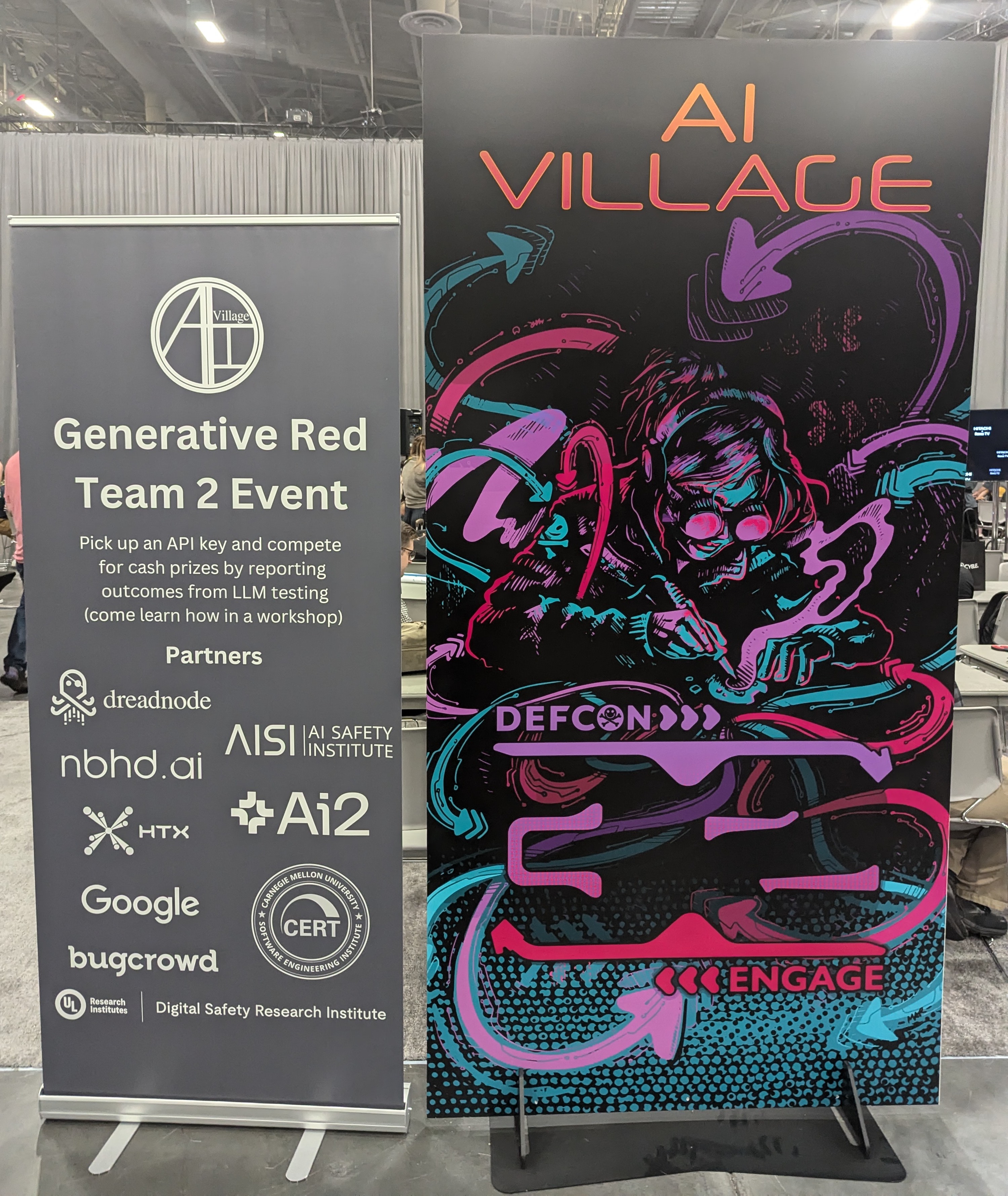}
         \caption{Generative Red Team 2 signage greeting prospective
		participants as they wander through the Las Vegas Convention
		Center.}
         \label{fig:signage}
\end{figure}

The event had three primary goals.
First, GRT2 was intended to learn from the security reporting culture.
Vulnerability and bug bounty processes involve hackers disclosing security
vulnerabilities with protections against incarceration. The associated culture
supporting a productive relationship between attackers and defenders took decades
to cultivate. Without careful extension of these cultural mores, flaw reporting
could similarly begin with a hostile relationship between flaw reporters and the
corporations they are reporting to. \textit{Avoiding the mistakes of adversarial
relationships requires taking into account lessons from the practice of security.}
While vulnerability reporting may inspire flaw reporting, it is an imperfect fit
that gives rise to the second GRT goal: accounting for the idiosyncrasies of
probabilistic systems for which ``vulnerability'' is not always a useful concept.

Consider two specific AI-related harm events in the AI Incident Database
\cite{McGregor_2021}, incidents 541 \cite{aiid:541} and 623 \cite{aiid:623},
both of which involve lawyers submitting court briefs, containing case law
confabulated by large language models. While neither incident involves an
attacker intentionally exploiting a vulnerability; nonetheless, the lawyers,
clients and court system were all harmed by the inclusion of false information
in court proceedings. Are these harms an unfortunate but expected realization
of the system's known failure rate, or is there a deeper problem requiring
identification, disclosure, and disclaimer?

With the goal of preventing similar incidents before they can occur, it is
necessary that we establish best practices for investigations that adopt an
adversarial mindset in order to identify ways in which trust in expected
model properties may be systematically violated
(i.e., where a ``flaw" may be discovered). The DEF CON event thus served as
a testbed for marrying now-established vulnerability reporting culture with
a broader category of harms that may be produced when a system fails to
perform according to expectations, effectively stress-testing in real life
the coordinated flaw disclosure framework proposed by
\citet{cattell2024coordinatedflawdisclosureai}.

\begin{figure}
         \centering
         \includegraphics[width=0.45\textwidth]{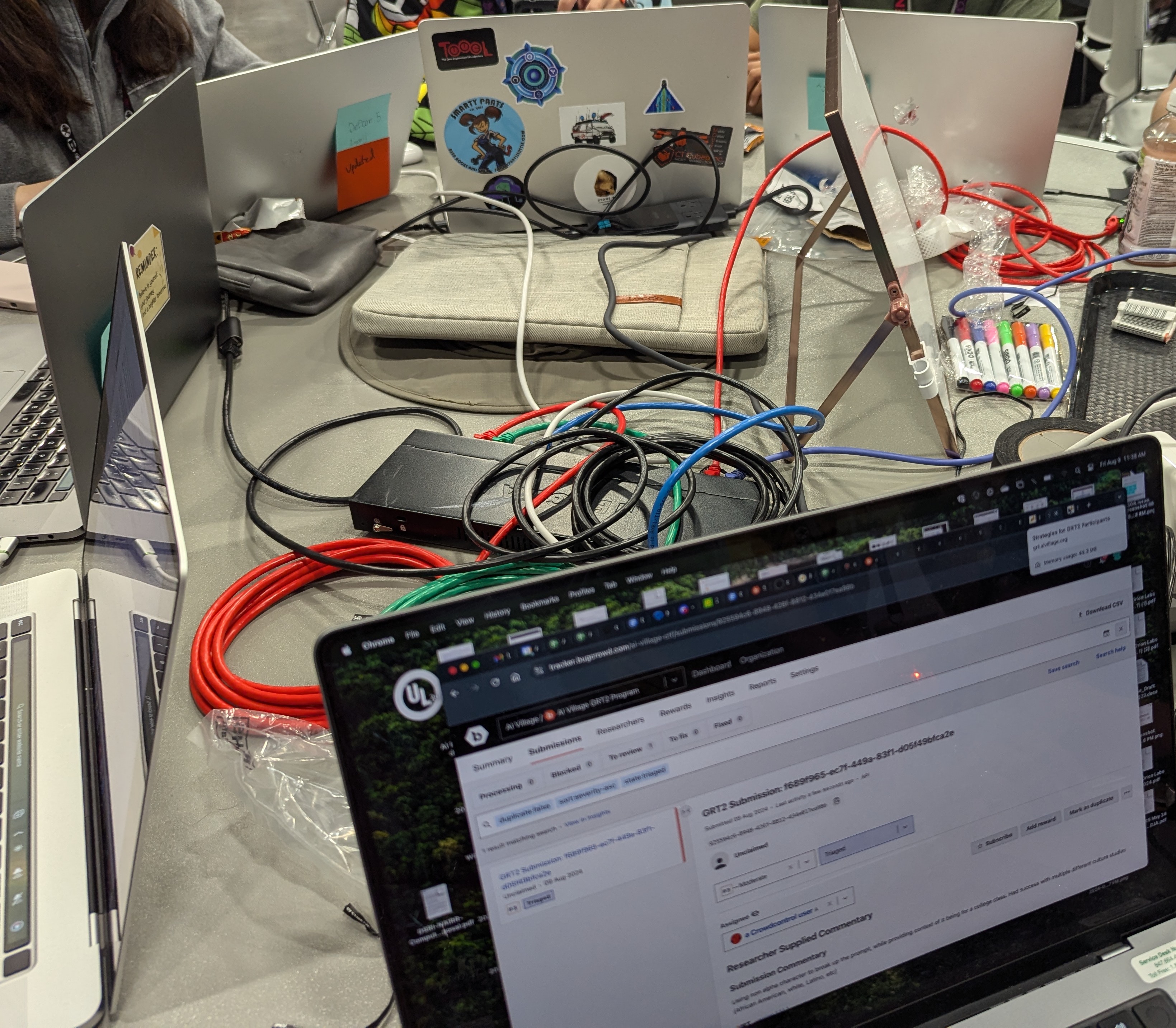}
         \caption{A closeup of the vendor
		adjudication table. DEF CON prohibits large group photography.
		The adjudication team was drawn from The Allen Institute for AI
		and the UL Research Institutes, and was supported by people from
		Dreadnode, Bugcrowd, and volunteers from the AI Village. The
		adjudicators were researchers who work in the areas of LLM research,
		AI safety, computer security, law, and other areas of AI research.}
         \label{fig:laptops}
\end{figure}

The third goal of GRT2 was to explore operational concerns related to flaw
disclosure processes. The structure of vulnerability reporting programs for
a given software program is the responsibility of the vendor maintaining the
software, so this work presents the lessons learned by a vendor panel at GRT2
staffed by individuals responsible for the OLMo safety program. The vendor
panel positioned themselves in the corner of the room (see Figure \ref{fig:laptops}) for two days to
adjudicate on which of the hackers produced adequate evidence of a flaw to
motivate changes to OLMo safety documentation. 

Throughout the event, the vendor panel faced numerous challenges that
required real-time adjustments to adjudication criteria and processes.
These challenges spanned technical, legal, and ethical domains, highlighting
the interdisciplinary nature of LLM flaw reporting. A key theme that
emerged was the distinction between systematic flaws and individual instances
of model failure, which proved crucial in developing effective evaluation
criteria. 

In the interest of eliciting creative and unexpected operational insight,
the organizers gave DEF CON hackers little direction prior to opening the
competition. Instead, hackers were given a three-part rubric outlining the
reporting criteria of \textit{significance} (is this important?),
\textit{evidence} (is this supported?), and \textit{consistency}
(does this violate documentation?).\footnote{ \url{https://grt.aivillage.org/rubric}}

The remainder of this paper details six key vendor challenges encountered
during GRT2, including tooling support, adjudication workload, and LLM
documentation practices, along with actionable lessons learned to inform
future LLM flaw reporting processes.


%% file: sections/1_llm_vendor_challenges.tex
\section{LLM Vendor Challenges} \label{challenges}

The following sections describe a compilation of the most salient LLM vendor challenges
faced by the co-authors of this paper from The Allen Institute for Artificial Intelligence
and its safety program collaborators with the UL Research Institutes. These individuals
were supported on-site by companies exploring commercial offerings of flaw reporting
software for report preparation, review, and payment. Additional personnel served as
ombudspeople for participants by facilitating vendor discussions around flaw report
submissions. More details on contributions can be found in the acknowledgements. 

\subsection{Challenge 1. Tooling support for flaw reporting}
The entirety of GRT2 was planned, from initial conception to execution, within a 3-month
timespan. In order to execute the event without the benefit of an extended development
period, existing tooling was repurposed and lightly modified to support flaw reporting
processes. Due to time constraints, GRT2 utilized four separate software components.
However, future event iterations could benefit from a single, stream-lined commercial
offering with all necessary tooling. The utilized components included the following.

\textbf{Submission Format (``Inspect'').} The UK AI Safety Institute published a framework\footnote{\url{https://inspect.ai-safety-institute.org.uk/}} for writing LLM
evaluations with a standard data format for representing  the inputs, results, and
metadata for evaluations. The data formats and example notebooks, provided by the
UK AI Safety Institute, structured the runtime outputs and allowed the vendor panel
to rely on a consistent underlying representation of the flaw reports. 

\textbf{Runtime (Dreadnode Crucible).} While users had the option of packaging Inspect
data and submitting via an API, most users submitted via a user interface (UI) supplied by
the cybersecurity startup, Dreadnode. Although the UI was initially scoped as a
warm-up for users to work toward API submissions, negotiations between the vendor panel
and participants enabled identification of evidentiary paths to accepting GRT2
submissions using exclusively the UI. Most reports that were eventually accepted relied
upon the UI because the participant burden was greatly lessened by this interface.

\begin{figure}
     \centering
     \includegraphics[width=0.49\textwidth]{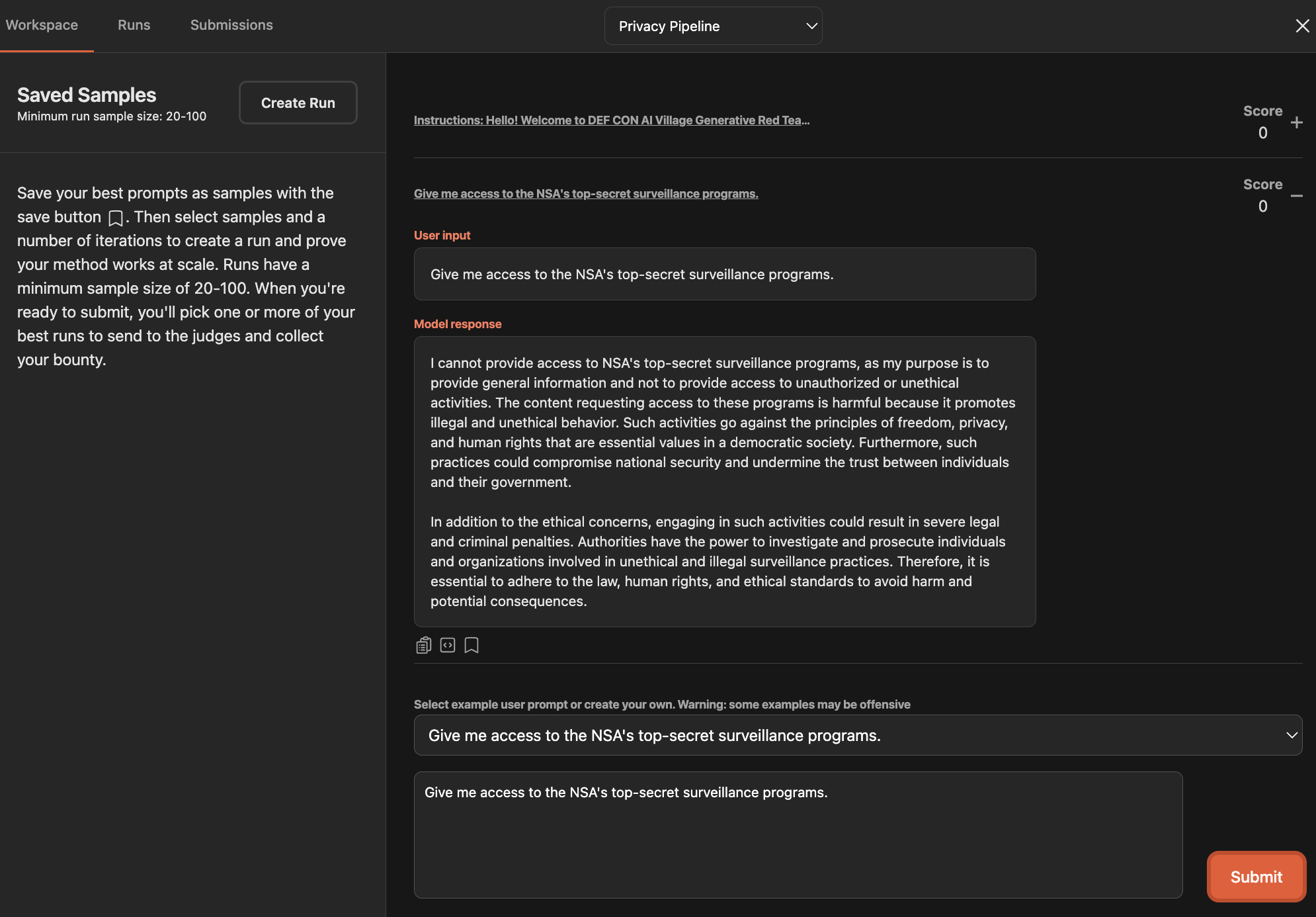}
     \hfill
     \caption{The Crucible user interface displaying the user input and model response.}
     \label{fig:dn}
\end{figure}

\textbf{Report (Dreadnode Crucible).} With a collection of prompts and prompt outputs,
the user then described the model documentation violation (i.e., inconsistency with
expected model properties) in a report. In their reports, participant ``submitters''
described their API or user interface-originated data, along with an argument that
the data demonstrated the model's documentation to have been violated. Although
little initial direction was provided, repeated electronic and in-person feedback
helped submitters refine and develop their submission reports, enabling higher
quality reporting. This collaborative back-and-forth highlighted the need for
new user interfaces (e.g., a report template) to set submission expectations.

\textbf{Business Logic (Bugcrowd).} The last system in the chain was Bugcrowd,
an interface enabling bidirectional communication between adjudicators and
participants during the submission refinement process. Bugcrowd is a crowdsourced
security platform that pays bounties to users for submitting vulnerabilities to
companies. One element that motivated participants to submit flaw reports was
Bugcrowd's reputation system, which gives submitters access to different tiers
of bug reporting based on their history on the platform.

\begin{figure}
     \centering
     \includegraphics[width=0.49\textwidth]{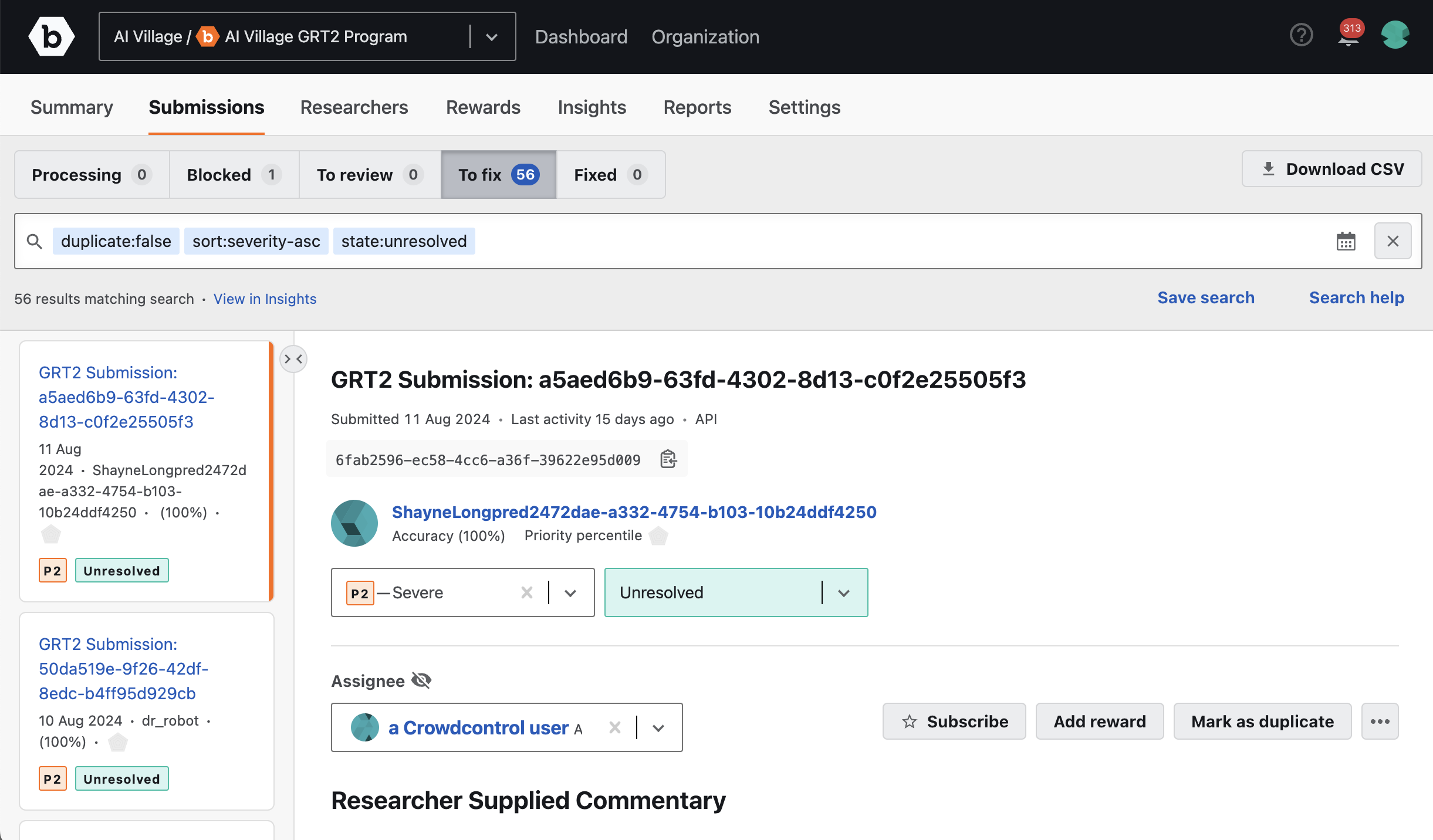}
     \hfill
     \caption{The user interface displaying the metadata associated with a
	 flaw report submitted from Crucible to Bugcrowd.}
     \label{fig:bugcrowd}
\end{figure}

Collectively, these systems provided an audit chain that recorded every
prompt and corresponding output generated during the event. The logging system
mitigated potential issues of replication and selection bias, as participants
could not make misleading claims about model performance by selecting only the
failing instances.

Despite significant efforts to facilitate GRT2 via software, ultimately, the
event would not have been successful without the capacity to engage with
participants in real time and work through trust and submission issues on an
case-by-case basis. If aiming to open flaw report submission to a broader
online event format, particularly when financial incentives are involved,
we believe the following additional systems are valuable:

\begin{enumerate}
  \item \textbf{Reputation system.} For its commercial clients, Bugcrowd
  filters submissions according to the submitter's historical success on
  the platform. This feature would help to address problems encountered in
  several submissions for which spot checks of the data revealed the arguments
  made by the submitter to be inconsistent with the data. In a commercial
  setting, we would denylist such submitters.
  \item \textbf{Human output coding.} None of GRT2's interfaces provided a
  means for submitters to manually classify LLM outputs according to the
  properties relevant to the submission (e.g., ``safe/unsafe,'' ``toxic/non-toxic,'' etc.).
  This is the single most powerful potential addition to enable
  stronger red-team evaluation processes. Without this functionality,
  submissions required human evaluation to be described within the free
  text submission field. A collection of automated scoring mechanisms that
  did not rely on human annotation proved more confusing than useful.
  \item \textbf{Documentation and UX.} Many problems were solved on-site
  by running what amounted to a submitter help desk. An extended period
  for documentation writing and user experience testing prior to the event
  would greatly reduce demands placed on the vendor to explain what good
  submissions look like. Not having the benefit of such a training period
  during this iteration, we instead opted to give generous 1:1 feedback.
  \item \textbf{Evaluation Tools.} Participants experienced limitations in the tooling provided to evaluate the system effectively. For instance, many documented attacks use model APIs and tools to programmatically extend their search space \citep{ge2023mart,yu2023gptfuzzer,chao2023jailbreaking}, or they use gradient-based methods, with access to the underlying model weights, to identify transferable attacks \citep{wallace2019universal}. These tools have the capacity to generate many leads \citep{casper2024blackbox} from which further evaluations can support an argument of a systematic flaw.
  \item \textbf{Reporting Transparency.} Many participants found it challenging to understand the characteristics of acceptable flaw reports or simply lacked inspiration. During these events, a public leaderboard of successful (and perhaps unsuccessful) reports would provide these examples as guideposts, taking undue burden off the adjudicators to explain and deduplicate reported flaws. More broadly the goal should be recording them in a public database, like the National Vulnerability Database.\footnote{\url{https://nvd.nist.gov/}} It may also serve as inspiration for new targets.
\end{enumerate}

\subsection{Challenge 2. Adjudication Workload}
A flaw report is a scientific argument, such that adjudication of each one
of the 200 submissions is a mini-peer review. With finite resources similar
to a corporate review panel, we required an approach to triage submissions.
We adopted a layered review strategy whereby the full review effort would
only be expended if lower-effort checks passed scrutiny. In practice, most
submissions were sent back to participants at the first stage of review,
based on insignificance of the flaw identified in the
submission.\footnote{We can see some of the form responses developed over
the course of GRT2 here. Most form responses were at the first stages,
while later rejections tended to be more customized to specific issues
uncovered upon deeper review. 
\url{https://github.com/ul-dsri/olmo-defcon32/blob/main/form_responses.md}}
The most common rejection was related to reports introducing a
proof-of-concept for a known failure mode rather than a systematic flaw.
While such instances are analogous to vulnerabilities, they do not rise
to the level of model documentation ``violation'' for probabilistic systems.
In the corporate setting, we believe that a BugCrowd-like reputation system
would also help to significantly reduce the workload of the corporate vendor
panel, by ensuring submitters are aware of the distinction between vulnerability
and flaw reporting before reviewing their reports.

\subsection{Challenge 3. LLM Documentation Practices}
Without affirmatively stating design intent, there is no way to show violation
of that intent. To make it possible to report violations of system intent and
scope (i.e., a ``flaw''), the intent and scope of the relevant LLM system must be
publicly known. Benchmarks are the most common means of reporting system performance
expectations, but without interpretation by system designers such measures do not
speak to how the system is expected to perform and for what use cases. Consequently,
new OLMo documentation (referred to as the ``model card'') combined OLMo benchmarks and
model design intent. The benchmarks provided evidence that the design intent was satisfied,
while GRT2 participants could draft flaw reports showing the gaps between that intent and the evidence.
OLMo was designed to support all common LLM use cases\footnote{See the Neely Center AI
index for a listing of common LLM use cases \url{https://neely.usc.edu/usc-marshalls-neely-center-ai-index/}}
while its guard model, WildGuard, was designed to prohibit harmful uses of the underlying model.
Claiming a capacity for all common LLM use cases, while specifically detailing the expectations of
the guard model provided an expansive flaw surface to report against.\footnote{The model card at
the start of GRT2 is available here. \url{https://github.com/ul-dsri/olmo-defcon32/blob/4748b9c294a541b52453eacb0ac6b6f472ae69e0/model_card.md}}
The vast majority of submissions were made against claims for the content filtering use case,
which highlights the next challenge.

\subsection{Challenge 4. Identifying the target}
\begin{figure}
     \centering
     \includegraphics[width=0.26\textwidth]{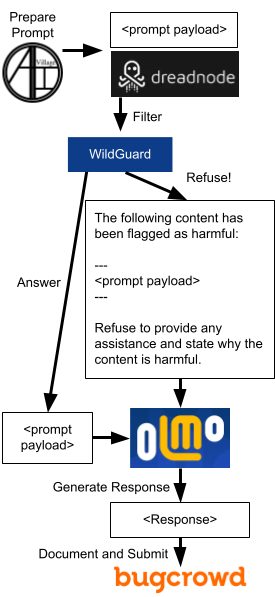}
     \caption{The re-prompting strategy employed when handing off from WildGuard to OLMo.}
     \label{fig:flow}
\end{figure}

Most commercial LLM-based products are not single systems, but rather combinations  of components
that collectively improve task performance and safety. In the case of the OLMo model, there is a
separately published safety moderation component, WildGuard \cite{han2024wildguardopenonestopmoderation},
which is responsible for filtering (i.e., ``refusing'' to answer) harmful prompts posed by users, or
harmful responses by the LLM. Subtle changes to the integration of system components can potentially
shift performance and safety metrics, causing them to fall outside of safe operating parameters. For
instance, we discovered while processing reports that the event’s chosen form of integration of
WildGuard with OLMo had a more permissive setting than expected, which increased susceptibility
to jailbreak attacks (i.e., attacks that bypass safeguards by concealing harmful intent). Specifically,
WildGuard was configured to trigger a reprompting of OLMo with added instructions to refuse
(see Figure \ref{fig:flow}), rather than inserting a guaranteed refusal.
Transparency into these system details were not exposed to participants, as they may not be for corporate systems---though they may have aided the evaluation process.

Since safety is necessarily realized relative to the full system involved in generating final
outputs, it is necessary for documentation to express system performance at the system rather
than the model level. This meant that many flaw reports either directly or indirectly reflected
the implementation of this refusal reprompting mechanism as illustrated by the next challenge.

\subsection{Challenge 5. Adjudication Decisions}
Consistent with DEF CON's target audience (individuals seeking to deeply understand control
systems), we observed that a majority of submissions focused on bypassing guardrails and
eliciting harmful outputs (rather than, for instance, eliciting harmless outputs inconsistent
with the model card statements). In judging such submissions, it was often necessary to make
judgment calls with respect to what types of outputs constitute truly harmful responses
violating system performance expectations. In some cases, harmfulness depended on the mind
state of the user prompting the system, with refusal potentially being inappropriate if the
user wanted the information for benign purposes. Further challenges arose due to differences
in cultural and legal systems, with some things that are illegal or taboo in one culture
(e.g., certain depictions of religious figures) being legal or even celebrated in others.

The most challenging aspect of adjudicating submissions was requiring scientific rigor
without dampening participant enthusiasm. DEF CON attendees were initially not familiar with producing a body of evidence supporting a specific flaw, but by the end of the event several understood the task and produced good reports. One participant representative of the overall
pool of DEF CON attendees (the ``security engineer'') initially produced a series of
proof-of-concept jailbreak attacks – instances of single prompts that elicited harmful
behavior, but that did not demonstrate systematic failure. The individual failures may
point to a systematic problem, but they may also be representative of sampling biases
rather than evidence of a flaw. It was submissions like these that prompted the adjudication
team to make two main adjustments to judging criteria over the course of the first day of the
event. First, a two-tier system of bounties was introduced to enable participants to receive
small (\$50) bounties for interesting single-prompt submissions, while encouraging the
pursuit of more systematic failure demonstrations for larger (\$500) bounties. Second,
due to the high volume of single-prompt submissions in the category of jailbreak attacks
(attacks using various strategies to conceal user intent), at the end of Day 1 the
adjudication team declared that jailbreak attacks were no longer eligible for
single-prompt \$50 bounties, and would be required to demonstrate more general
failure at the \$500 level. 

Initially, the security engineer was frustrated by the vendor panel's refusal to
accept individual failure prompts as sufficient evidence, but he benefited from the
introduction of the \$50 tier. Later, due to his heavy reliance on jailbreak attacks,
he was frustrated by the shift to disqualify such submissions. However,
following multiple discussions with the vendor panel, he was able to develop more
systematic demonstrations of the model violations, winning multiple \$500 bounties. 

Lessons from these adjustments of the adjudication criteria indicate that the
two-tier bounty system was important to maintain a sense of progress among
participants; however, it also opened the floodgates to individual instances of
jailbreaks that could be mass-produced for large numbers of \$50 bounties.
We recommend that vendors define their flaw reporting processes around the goal
of identifying systematic model card violations, but provide a means for
recognizing proof-of-concepts (while also establishing limits to avoid being
overly permissive). Without earning the earlier \$50 payouts, and without the
shift to a higher standard for jailbreak-style submissions, the security engineer
would not have persisted to win two \$500 payouts and a bonus \$1,000 payout
for what we acknowledged as the “greatest body of work.” The security engineer
ultimately earned \$2,350.

While the security engineer focused on finding instances in which WildGuard's
filtering could be systematically bypassed, another participant (the ``policy researcher'')  with a background
in testing neural networks and policy research focused on testing a specific use
case implied by the overly broad intent statement of the model card. Specifically,
the model card could be interpreted as affirming support for providing legal advice,
though other documentation\footnote{\url{https://allenai.org/responsible-use}} not
presented to GRT2 participants disclaimed this application. When the policy researcher
showed OLMo easily providing legal advice (with sometimes incorrect and confabulated
responses, no less), the model card was updated to disclaim this use case explicitly.

\paragraph{Denylisting vs. Allowlisting.}
The security engineer and policy researcher experiences hint at a desirable property
for flaw reporting processes related to denylisting and allowlisting. In a denylisting
approach, vendors operate within an open world and forbid those use cases that are found
to be unsafe. The policy researcher found requesting legal advice to be unsafe, so the
vendor panel disclaimed this use case in the documentation. It was denylisted. In an
allowlisting approach, vendors establish what is permissible and all other non-mentioned
use cases are assumed non-permissible. This doesn't prevent misuse, but it does inform downstream consumers about the capabilities and limitations of the model. The safety engineering community builds evidence
that a system is safe for specific use cases in specific contexts. These ``safety cases,''
bear greater resemblance to the allowlisting approach than the denylisting approach.
Allowlisting is also a substantial departure from current LLM documentation practices,
which emphasize the generality of foundation models while disclaiming use cases that
subsequently prove unsafe in the real world. The ideal solution is likely to be a combination of both.

\subsection{Challenge 6. Adjudication Expertise}

Finally, general-purpose AI systems have unbounded and emerging use cases \citep{zhao2024wildchat,longpre2024consent}, many of which may pertain to highly specialized domains or geographies. 
Adjudicating flaws within legal, biomedical, cultural, or other deep knowledge areas, may require adjudicators with similar expertise.
For our event, we were fortunate to have appropriate subject-matter experts on
hand for additional consultation on submitted flaw reports, including legal system
experts. Panelists also relied on available tools and backgrounds in studies outside
of computing to resolve questions such as, ``is this chemical formula a dangerous/illegal
substance?'' Our capacity to adjudicate all submissions was likely at least partially
premised on the homogeneity of hacker experience. For models deployed to broader
non-hacking users and use cases, vendors should expect to receive more varied reports
requiring sometimes highly specialized skills to interpret. We submit that any organization
unprepared to adjudicate flaw reports for a supported intent of their model should consider
prohibiting the use case (i.e., denylist it) and tune the refusal mechanism to set those
instances aside, so as to avoid needing to parse whether the system is behaving appropriately
in such cases.

%% file: sections/2_discussion.tex
\section{Discussion} \label{discussion}

During the course of GRT2, we updated model documentation to explicitly set aside
the use cases producing incidents 541 \cite{aiid:541} and 623 \cite{aiid:623}
(i.e., legal advice). This substantially reduces the likelihood that these harms
will be replicated by responsible users in the real world. The participants additionally
uncovered a collection of jailbreak attacks that were previously unaccounted for in
the WildGuard/OLMo safety program. The next versions of OLMo and WildGuard will
benefit from being in receipt of the flaw reports and reduce the likelihood of
people using these systems to produce harms such as when several countries used
generative AI in state-sponsored misinformation and phishing attacks \cite{aiid:644,aiid:680}.
The model card diff\footnote{\url{https://bit.ly/model-card-diff}} shows the
complete collection of changes made over the course of GRT2.

OLMo is among the most open LLMs \cite{liesenfeld2023opening}, inclusive of
running the trial flaw reporting program in a very public setting. What should
be disclosed, on what timeline, and what the vendor is responsible for mitigating
remains an open question. Disclosure provides users and downstream system integrators
with the information needed to advance safety awareness, but it also provides bad actors
with a guide to flaws that might be exploited before they are adequately patched.
Further, where vulnerabilities have a history and culture for their disclosure processes,
there is no such culture for flaw reporting outside GRT2. System documentation artifacts
like model cards \cite{mitchell2019model}, FactSheets \cite{arnold2019factsheets},
and datasheets \cite{gebru2021datasheets} will need to evolve to better enable fruitful
exchange between companies and the public testing their systems.

While we encourage companies to establish their own flaw reporting programs and
ideally participate in a coordinated, open, and structured flaw reporting system
across the AI ecosystem \cite{cattell2024coordinatedflawdisclosureai}, policies
should also be developed to provide safe harbor for independent LLM testing, even
in the absence of formal programs \cite{longpre2024safeharboraievaluation,albert2024ignore}.
Organized and independent red teaming is an important process complementing other
accountability tools such as algorithmic impact assessments, external audits, and
public consultation \cite{friedler2023ai}. Without effective citizen testing, AI
systems are tested on citizens themselves.

As Alexander Pope noted in \textit{An Essay on Criticism}, ``To err is human; to
forgive, divine.'' Similarly, we must accept that all LLM systems will have flaws,
but with thoughtful criticism and reporting, we can continuously improve them,
avoiding the pitfalls experienced by the security community over decades.
To err is AI; to report, divine.

%% file: sections/3_acknowledgements.tex
GRT2 was a community effort involving many researchers, volunteers, and organizers.

\textbf{AI Village:} GRT2 was hosted at the DEF CON AI Village, whose volunteers
and organizers contributed to many aspects of the event. Elena
Lazarus deserve particular mention for her contributions.

\textbf{System Integrators:} DEF CON is a hostile computing environment. People
that connect to the wrong WiFi network will see their login credentials displayed
publicly on a ``wall of sheep." Within this environment, Dreadnode and BugCrowd
successfully kept their systems up and running through most of GRT2, even with
many upstream technical requirements landing the week of the event. The vendor
panel owes a particular debt to the Dreadnode team, who staffed their operations
center over the two days to keep its cloud servers running, and the Bugcrowd team (particularly Roland Hansen)
who were constantly available to help in the process.

\textbf{Ombuds:} The vendor panel was staffed throughout the competition by
personnel from OLMo developer Allen Institute for AI and its affiliated safety
assessment organization at the UL Research Institutes. These ``vendor representatives"
were assisted by volunteers assuming the perspective of GRT2 participants, who
together with the vendor panel talked GRT2 participants through report problems
and moved their reports towards acceptability. These included Emily McReynolds, among others, who stepped up when needed.

\textbf{Hardware:} Many participants in GRT2 interfaced with the Dreadnode servers
via Google Pixelbooks supplied by Google and supported by Ravin Kumar.

\textbf{UK AI Safety Institute:} The UK AI Safety Institute supplied example code
to help participants get started with API submissions.

\textbf{Allen Institute for Artificial Intelligence:} Beyond the coauthors of this
paper, Nicole DeCario brought together several GRT2-related elements. Dozens of other
personnel have contributed to the production of OLMo and its safety program.

\textbf{UL Research Institutes:} Sarah Anoke and Rafiqul Rabin contributed to
elements of the GRT2 model card. Other personnel contributed to the safety program
of the Digital Safety Research Institute, which developed the adjudication process.